\documentclass[a4paper,11pt]{article}
\usepackage{pos}

\title{Advanced techniques of searching for flares of ultra-high-energy photons from point sources}
\ShortTitle{Advanced techniques of searching for flares of ultra-high-energy photons}

\author*[a]{Jaroslaw Stasielak}
\author[a]{Chaitanya Priyadarshi}
\author[a]{Dariusz Góra}
\author[a]{Nataliia Borodai}
\author[b]{Marcus Niechciol}
\author[a]{Jan Pękala}

\affiliation[a]{Institute of Nuclear Physics Polish Academy of Sciences, PL-31342 Krakow, Poland}

\affiliation[b]{Department of Physics, University of Siegen, Siegen, Germany}

\emailAdd{jaroslaw.stasielak@ifj.edu.pl}
\emailAdd{chaitanya.priyadarshi@ifj.edu.pl}
\emailAdd{dariusz.gora@ifj.edu.pl}
\emailAdd{nataliia.borodai@ifj.edu.pl}
\emailAdd{niechciol@physik.uni-siegen.de}
\emailAdd{jan.pekala@ifj.edu.pl}

\abstract{Astrophysical flares are one of the possible prominent source classes of ultra-high-energy (UHE, $E > 10^{17}$ eV) cosmic rays, which can be detected by recording clusters of extensive air showers in arrays of detectors. The search for sources of neutral particles offers distinct advantages over searching for sources of charged particles, as the former traverse cosmic distances undeflected by magnetic fields. While no cosmic-ray photons exceeding $10^{17}$ eV have been definitively detected, identifying the clustering of events in cosmic-ray data would provide compelling evidence for their existence.
We compare two analysis methods for detecting direction-time clustering in UHE extensive air showers: an approach in which one examines multiplets, and the stacking method, in which one analyzes sets of doublets that are not necessarily consecutive, thus making it sensitive to multiple flares. Both techniques combine time-clustering algorithms with unbinned likelihood study. Background events (initiated by hadrons) can be more efficiently distinguished from photon-induced events (signals) by using a photon tag that employs probability distribution functions to classify each event as more likely to be initiated by either a photon or a hadron. We demonstrate that these methods can effectively distinguish between events initiated by photons and those initiated by hadrons (background), and can accurately reproduce both the number of photon events within flares and their duration. We calculate the discovery potentials, i.e., the number of events required to identify a photon flare. The methods discussed can be used to search for cosmic ray sources and/or improve limits on the fluxes of UHE photons.}

\ConferenceLogo{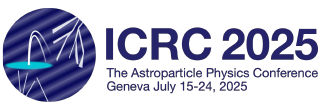}

\FullConference{39th International Cosmic Ray Conference (ICRC2025)\\
 15–24 July 2025\\
Geneva, Switzerland\\}

\begin{document}
\maketitle

\section{Introduction}

Ultra-high-energy cosmic rays (UHECRs; $E > 10^{17}$ eV) are the most energetic particles observed in the universe, and identifying their sources remains a fundamental challenge in astrophysics. Among various cosmic phenomena, active galactic nuclei flares and gamma-ray bursts are particularly promising source candidates. If a fraction of UHECRs originates from astrophysical flaring events (point sources), their arrival at Earth would be expected to occur in temporally and directionally clustered groups. Such direction-time clustering should be detectable for neutral particles such as ultra-high-energy (UHE) photons, which travel in straight paths from their sources to the detector, unlike charged cosmic rays that are deflected by cosmic magnetic fields. This property makes UHE photons ideal for source identification, as they point directly back to their origins, with accuracy limited only by the angular resolution of detection systems. Although no cosmic-ray photons with energies above $10^{17}$ eV have yet been conclusively observed, the detection of clustered events in cosmic-ray data would provide compelling evidence for their existence. This motivates the development of sensitive methods for identifying such clustering patterns.

In this paper, we present two approaches for finding the direction-time clusters in UHECR (air-shower) data \citep{BRAUN2010175,GORA2011201}. Both techniques integrate time-clustering algorithms with unbinned likelihood analysis and have been previously applied to investigate correlations between neutrino arrival directions detected by the IceCube Neutrino Observatory and the positions of their potential astrophysical sources. We enhance these methods by incorporating a discriminating factor, referred to as the photon tag, which helps separate signal events (photon-initiated air showers) from background events (hadron-initiated air showers), and adapt them for use with cosmic ray data.
Our analysis demonstrates that these advanced methods can effectively identify UHE photon flares. We highlight the robustness of the approach, as it requires only a few signal events for reliable flare detection. These methods may prove useful in identifying cosmic ray sources and in establishing stronger limits on the fluxes of UHE photons.

\section{Direction-time clustering analyses} \label{sec:description}

We propose two enhanced analysis methods for detecting direction-time clustering in UHE extensive air showers. One approach focuses on the examination of multiplets (multiple consecutive events), so-called the Braun method \citep{BRAUN2008299,BRAUN2010175},  
whereas the other one utilizes the stacking method \citep{GORA2011201,stasielak2024,stasielak2025b}, in which one analyzes collections of doublets (two events occurring one after another) that need not be temporally adjacent, thereby enabling detection of multiple flares with arbitrary temporal profiles. Both techniques utilize unbinned likelihood analysis. In this framework, identifying a point source of photons (corresponding to a flaring event) requires detecting a statistically significant cluster of direction-time correlated events from a specific sky direction. We begin our description of the methods with the definition of the likelihood function $\mathcal{L}$, which is central to our analyses.

The likelihood is defined as
\begin{equation}
\mathcal{L}(n, \Delta t_j) = \prod_{i=1}^{N} \Bigg(\frac{n}{N}s_{i} + (1 - \frac{n}{N}){b}_i\Bigg) \rm{,} \label{eq:like}
\end{equation}
where we combine the signal ($s_i$) and background ($b_i$) probability distribution functions (PDFs) for all observed events (indexed by $i$). Here, $N$ denotes the total number of events in the considered data sample, $n$ is the assumed number of signal events, and $\Delta t_j$ is the time window of the considered multiplet or doublet.  A background-only likelihood, representing the absence of signal events, simplifies to $ \mathcal{L}(0, \Delta t_j)=\prod_{i=1}^{N} b_{i}$.

For each event $i$, both the signal PDF ($s_i$) and background PDF ($b_i$) are factorized into directional, temporal, and photon tag components: 
\begin{equation}
s_i=s_i^{\mathrm{direction}}s_i^{\mathrm{time}}s_i^{\mathrm{tag}} , \hspace{1cm} b_i=b_i^{\mathrm{direction}}b_i^{\mathrm{time}}b_i^{\mathrm{tag}}, 
\end{equation}
where the photon tag components $s_i^{\mathrm{tag}}$ and $b_i^{\mathrm{tag}}$ are due to an additional discriminating factor, referred to as the photon tag, which employs PDFs to classify each event as more likely to be either a signal (photon-initiated air shower) or background (hadron-initiated event). A detailed description of these factors is postponed to Section \ref{ph-tag}.
The directional component of the signal PDF ($s_i^{\mathrm{direction}}$) is modeled as a two-dimensional Gaussian distribution
$s_i^{\mathrm{direction}} = \frac{1}{2\pi \sigma_i^2} \exp \left(-\frac{\left|\vec{r}_i - \vec{r}_s \right|^2}{2\sigma_i^2}\right)$,
where $\vec{r}_i$ is the reconstructed direction of the event, $\vec{r}_s$ is the assumed source direction, and $\sigma_i$ is the angular uncertainty associated with event $i$. 
For each tested time window $\Delta t_j = t^{\rm{max}}_j - t^{\mathrm{min}}_j$, the temporal signal PDF ($s_i^{\mathrm{time}}$) is defined as 
$s_i^{\mathrm{time}}=\frac{H\left ( t_{j}^{\mathrm{max}}-t_i \right )H\left ( t_i-t_{j}^{\mathrm{min}} \right )}{\Delta t_{j}}
\label{sig_time}$,
where $H$ represents the Heaviside step function and $t_i$ is the arrival time of the event. 
This formulation ensures that only events within the specified time window $\Delta t_j$ contribute to the signal, with $s_i=0$ for all events outside this interval. 

The directional component of the background PDF is defined based on the total solid angle $\Delta \Omega$ subtended by the analyzed sky region, and the temporal component is determined by the detector uptime $\Delta T_{\mathrm{data}}$. These are based on the assumption of a uniform distribution of the background in both domains. This yields the following background PDFs: $b_i^{\mathrm{direction}} = 1 / \Delta \Omega$ and $b_i^{\mathrm{time}}=1 / \Delta T_{\mathrm{data}}$.

Additional observables, such as energy information, could also be incorporated into the PDF framework to improve the ability to distinguish genuine UHE photon clustering from background fluctuations. However, this is omitted in the present analysis.

The test statistic ($TS$) is calculated as the logarithm of the likelihood ratio:
\begin{equation}
    TS(n, \Delta t_j) = -2 \ln \Bigg[\dfrac{\mathcal{L}(0,\Delta t_j)}{\mathcal{L}(n,\Delta t_j)} \Bigg] = 2  \sum_{i=1}^{N} \ln  \Bigg[ \frac{n}{N} \Big(\frac{s_{i}}{b_{i}} - 1 \Big) + 1\Bigg] \rm{.} \label{eq:ts_base}
\end{equation}
Using the fact that by construction, $s_i$ is non-zero only within the considered time window $\Delta t_j$, the $TS$ can be simplified to
\begin{equation}
TS(n, \Delta t_j) = 2 \sum_{i}^{N_{in,j}}\ln \Bigg(\frac{n}{N} \bigg(\frac{s_i}{b_i} - 1 \bigg) + 1 \Bigg) + 2 \cdot (N-N_{in,j})\cdot\ln \bigg(1 - \frac{n}{N}\bigg) \rm{,} \label{eq:ts_final}
\end{equation}
where the summation runs over all $N_{in,j}$ events within the $\Delta t_j$ time window. This form of the $TS$ yields approximately a tenfold improvement in computational performance, particularly in the case of the computationally expensive multiplet (Braun) method. 

The overall procedure for the multiplet (Braun) method is as follows. We identify all consecutive multiplets (doublets, triplets, etc.) in the data and calculate the corresponding test statistic $TS$, which is then maximized with respect to $n$ to obtain  $TS(n_s, \Delta t_j)$ where $n=n_s$ is the maximizing value. The multiplet yielding the highest maximized $TS$ value is then selected. Its time width is taken as the estimator of the flare duration $\Delta T$, while the corresponding $n_s$ value provides the estimate of the number of signal events. The statistical significance of the results is assessed by comparing the value of the maximized $TS$ with a background distribution of maximum test statistic values from randomly scrambled maps mimicking the original dataset.

For the stacking method, the procedure is similar but includes three 
steps. First, events are pre-selected based on the ratio $s_i^{\mathrm{direction}}s_i^{\mathrm{tag}} / \left( b_i^{\mathrm{direction}}b_i^{\mathrm{tag}} \right)> \rm{S/B}$, where $\rm{S/B}$ is an adjustable signal-to-background threshold. Second, we construct doublets using consecutive pre-selected events and maximize the corresponding test statistic $TS$ for each doublet with the addition of a marginalization term $-2 \ln \frac{ \Delta T_{\rm{data}} }{ \Delta t_j}$. All doublets are ranked according to these obtained values, which are also used as their weights in the following stacking procedure. Third, we apply the stacking procedure, combining progressively more doublets starting with the highest-ranked one. For each such set consisting of m doublets, we calculate the combined test statistic $\widetilde{TS} (m)$ and maximize it. The final combination of doublets is the one yielding the highest maximized value $\widetilde{TS} (M_{\mathrm{opt}})$. The stacking method is described in more detail in \citep{stasielak2024,stasielak2025b}.

\vspace{-0.3cm}

\section[S4 photon tag]{$S_4$ photon tag}

\label{ph-tag}

\begin{figure}[t]
\begin{center}
\vspace{-0.7cm}
\includegraphics[width=0.4\textwidth]{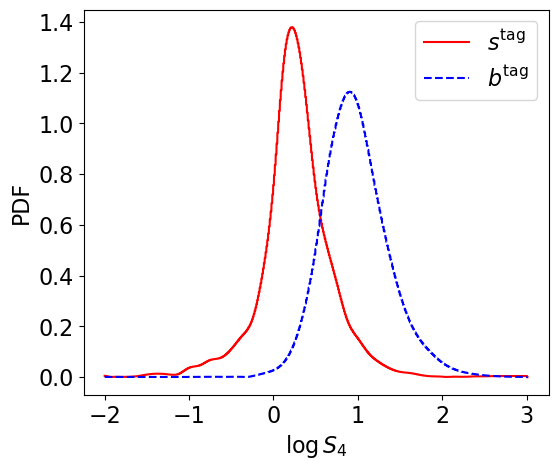}
\end{center}
\vspace{-0.7cm} 
\captionsetup{width=1\textwidth}
\caption{
Probability distribution functions 
of the variable $\log S_4$, derived from simulations of extensive air showers initiated by photons (considered as the signal, shown in red) and protons (considered as the background, shown in dashed blue) \citep{universe8110579}. These functions are used as the $S_4$ photon tag.
}
\vspace{-0.3cm} 
\label{fig:geo1}
\end{figure}

To improve the capability of the methods to distinguish between photon and background events, we employ a specially designed observable that helps in this separation and use it to construct a photon identification tag. The most natural choice for this purpose is the depth of the shower maximum, $X_{\rm{max}}$, which is directly measured by fluorescence or radio detectors
in large air shower observatories. 
However, currently such measurements are available only for a limited subset of recorded events, and to extend this data sample, we have to wait for the accumulation of data from the radio detectors with their 100$\%$ duty cycle.
To enable analysis over a larger data sample, we instead adopt the $S_b$ variable, a commonly used observable for differentiating between photon-induced and hadron-induced showers. It can be derived solely from measurements of surface air shower detector array and is defined as $S_b = \sum_k S_k \left(R_k /1000 \hspace{0.1cm} \rm{m} \right)^b$ \citep{ROS2011140}, where the sum runs over all surface detectors recording non-zero signals $S_k$, and $R_k$ is the distance of the $k$-th detector from the shower axis. While the parameter $b$ is somewhat flexible, we use $b=4$, which is a typical choice for photon searches used at the Pierre Auger Observatory\citep{universe8110579}. 
By construction, $S_4$ is sensitive to the lateral distribution.

The $S_4$ photon tag is constructed based on the PDFs of the $\log S_4$ variable for photon-initiated (signal; $s^{\rm{tag}}$) and proton-initiated (background; $b^{\rm{tag}}$) showers, derived from simulations \citep{universe8110579}, see the red and blue lines in Figure \ref{fig:geo1}, respectively. 
The PDFs of $\log S_4$ assess whether events are more likely to be photon-initiated or background; as we can see, this separation is not perfect. As the photon tag, we use $s_i^{\mathrm{tag}}= s^{\mathrm{tag}}(\log S_4^i)$ and $b_i^{\mathrm{tag}}= b^{\mathrm{tag}}(\log S_4^i)$, where $S^i_{4}$ represents the $S_4$ value for the $i$-th event. Alternatively, we can use another variable providing better discrimination; see, for example, the investigation of the $P_{\rm{tail}}$ variable in \citep{natalia}. 
To assess the application of the photon tag, we compare our results with a baseline case where no photon tag is applied. In this scenario, we set $s_i^{\mathrm{tag}} = b_i^{\mathrm{tag}} =1$ for all events.

\begin{figure}[t]
\begin{center}
\vspace{-0.5cm}
\includegraphics[width=1.00\textwidth]{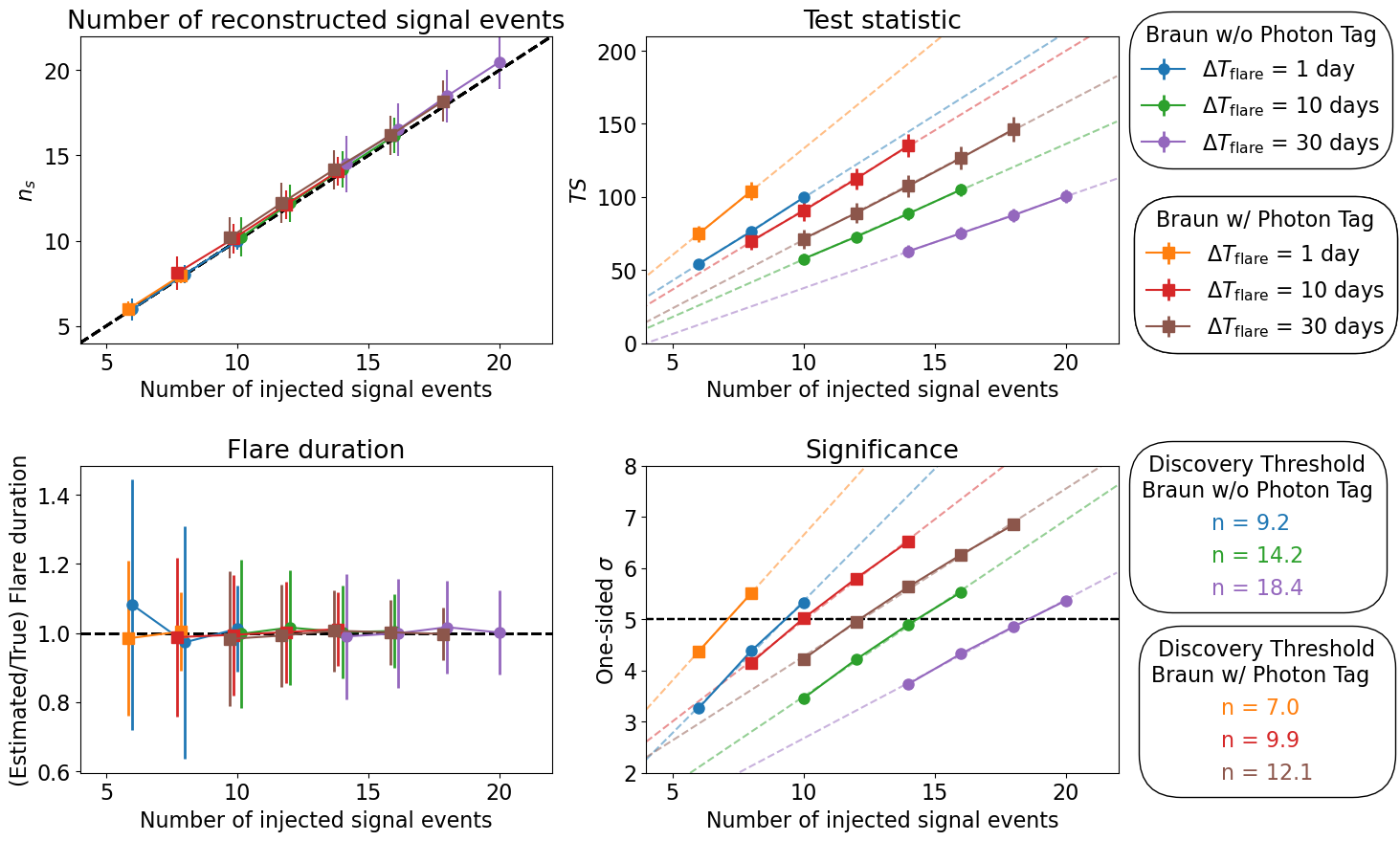}
\end{center}
\vspace{-0.3cm}
\captionsetup{width=1\textwidth}
\caption{
Plots showing the results obtained for the multiplet (Braun \citep{BRAUN2010175}) method with (square points) and without (circular points) application of the $S_4$ photon tag. 
The analyses were performed for flares of three different lengths: 1, 10, and 30 days. The plots on the left show the estimator $n_s$ (top) and the fraction of reconstructed to true flare duration for different numbers of injected signal events (bottom). 
The shown values (points) and their error bars are calculated as the mean and the width of the corresponding distribution obtained from multiple maps, respectively.
The plots for the different flare durations are shifted slightly to avoid the visual overlap and facilitate easier comparison.
The plots on the right show the $TS$ (top) and the statistical confidence level of the obtained results (bottom). 
The bottom right plot also shows the minimum number of events required to detect direction-time clusters at the $5\sigma$ confidence level, which is called the discovery threshold (see Section \ref{sec:discovery} for its exact definition). Discovery thresholds are obtained from the crossing between colored lines representing different simulated cases and the horizontal dashed black line (5$\sigma$ confidence level). The ranges of injected signal events are chosen differently for each analysis to capture the corresponding discovery threshold values, which are shown in the plot legend.
}
\vspace{-0.3cm}
\label{fig:comp_braun}
\end{figure}

\vspace{-0.1cm}

\section{Monte Carlo tests}

To evaluate the capability of the proposed methods to detect flares and accurately determine both the number of events and the flare durations, we performed Monte Carlo simulations. For this purpose, we generated numerous sky maps containing signal and background events. A total of 595 background events were uniformly distributed across a 12$^{\circ}$ x 12$^{\circ}$ sky region centered on the source, spanning an observation period of $\Delta T_{\rm{data}}=3150$ days. This setup mimics a subsample of Auger data used to evaluate different analysis methods. Signal events followed a Gaussian spatial distribution centered on the source, with arrival times uniformly distributed within the flare duration $\Delta T_{\mathrm{flare}}$. The start time of each flare was randomized.
For each map, we obtain the estimator $n_s$ of the number of injected signal events $N_s$, the estimator $\Delta T$ of the flare or flares duration $\Delta T_{\mathrm{flare}}$, and the maximum value of the $TS$ for the multiplet (Braun) method or of the $\widetilde{TS} (M_{\mathrm{opt}})$ for the stacking method. The test statistic is then used to determine the statistical significance of the result.

\begin{figure*}[t]
\vspace{-0.8cm} 
\begin{center} 
\includegraphics[height=4.9cm,trim={0cm 0cm 0cm 8cm},clip]{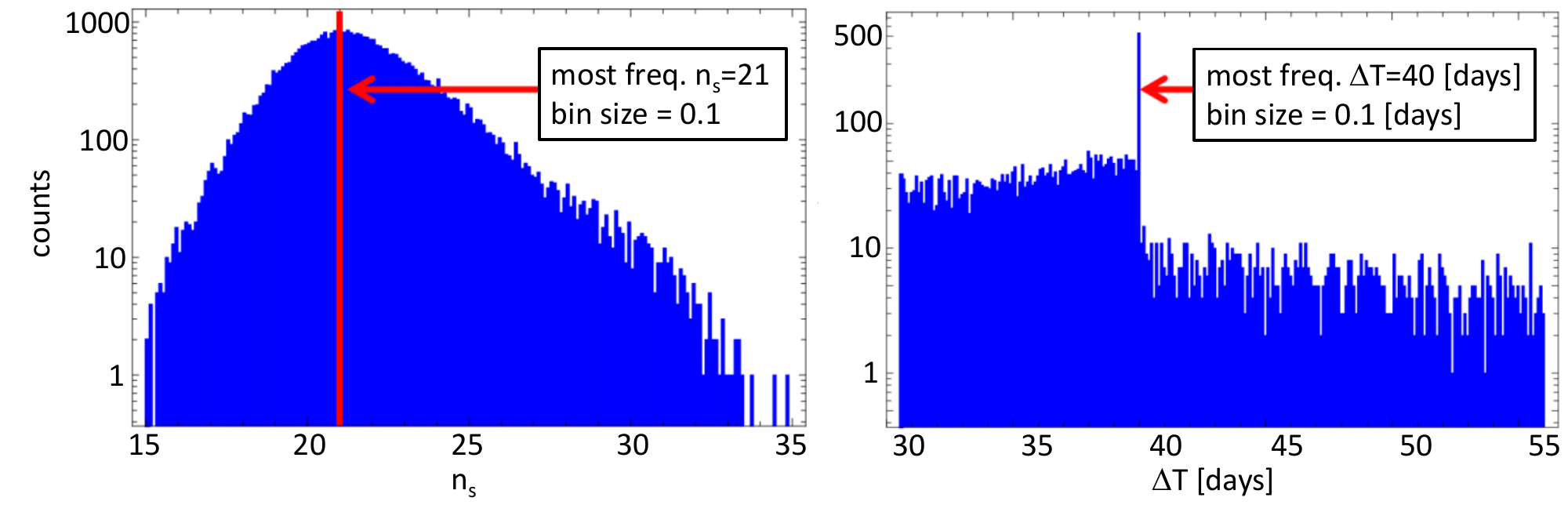}
\vspace{-0.4cm} 
\caption{
Distributions of $n_s$ (left) and $\Delta T$ (right) obtained from Monte Carlo simulations for the stacking method, applied to a triple-flare scenario (comprising three distinct flares of ten, ten, and twenty days duration) with $N_s = 20$ injected signal events. Each plot features a box highlighting the most common value, the peak of the distribution. 
The parameters of the flares are accurately reconstructed: the number of signal events is estimated as $n_s = 21$, and the combined flare duration is $\Delta T = 40$ days.
}
\vspace{-0.6cm}
\label{results}
\end{center}
\end{figure*}

Figure \ref{fig:comp_braun} shows the comparison of results obtained for the multiplet (Braun \citep{BRAUN2010175}) method, with and without application of the $S_4$ photon tag. The plots on the left highlight the effectiveness of the method in accurately retrieving the number of signal events and the flare duration. As we can see, the number of injected signal events is reconstructed with very small error, while the flare duration is typically determined with an accuracy better than 20$\%$ for the case with the photon tag. 
The plots on the right show the obtained $TS$ and the statistical confidence level, highlighting the robustness in detecting the flares. 
Note the higher $\rm{TS}$ values when the photon tag is applied, indicating that flares are more easily detected with its application. This is further supported by the bottom right plot, which shows the minimum number of events required to detect direction-time clusters at the $5\sigma$ confidence level -- referred to as the discovery threshold, as described in Section \ref{sec:discovery}.
The results clearly demonstrate that discovery thresholds are substantially reduced when using the photon tag.
All results presented above were obtained using a Python package named \texttt{UHECluster} (see \href{https://gitlab.com/uhecluster/uhecluster}{https://gitlab.com/uhecluster/uhecluster}), developed to provide a uniform, modular, script-based platform for running all the algorithms discussed in this work.

As for the stacking method, we present a single example of the $n_s$ and $\Delta T$ distributions obtained for a triple-flare consisting of one 20-day flare and two 10-day flares, with a total of 20 signal events (see Figure \ref{results}). The results demonstrate that this method can successfully reconstruct multiple flares.

\section{Discovery potential of the methods}\label{sec:discovery}

The efficiency of the method can be quantified by the minimum number of signal events required to detect a photon flare through detecting direction-time clustering of air shower events. It can be measured by a commonly used discovery threshold, defined as the number of signal events required to achieve a p-value less than $2.87 \times 10^{-7}$ (one-sided $5\sigma$) in $50\%$ of the trials (i.e. scrambled maps) (see \citep{stasielak2024,stasielak2025b} for more details). To calculate this threshold, we conducted extensive Monte Carlo simulations using scrambled sky maps to derive the $TS$ distribution for the background-only simulations. An exponential function was fitted to the distribution's tail to identify the $TS$ value corresponding to the $5\sigma$ significance level. We then examined how the $TS$ distribution evolves as we inject varying numbers of signal events into the background data. The discovery threshold is defined as the number of injected signal events at which the median $TS$ value reaches the pre-determined $5\sigma$ background threshold.

\begin{figure*}[t!]
\vspace{-0.8cm} 
\begin{center} 
\includegraphics[width=1.0\textwidth,angle=0]{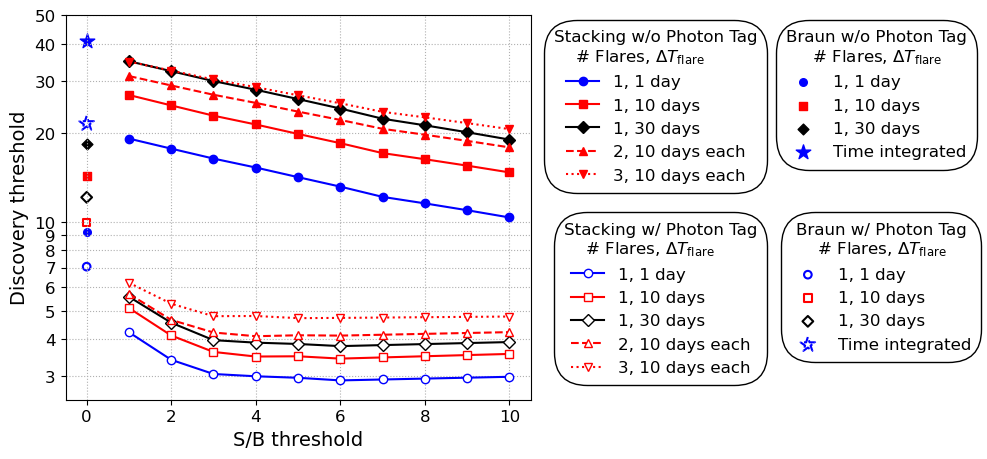}
\end{center}
\vspace{-0.5cm} 
\captionsetup{width=1\textwidth}
\caption{
Discovery thresholds for the stacking method, both with and without using the $S_4$ photon tag, are shown as a function of the $\rm{S/B}$ threshold. Results are presented for both single and multiple flares of varying lengths. 
These results are compared with the multiplet (Braun) method \citep{BRAUN2010175} and the time-integrated method \citep{BRAUN2008299} (star-shaped points), which employs direction-only clustering without time information, both without (filled markers) and with (hollow markers) application of the $S_4$ photon tag. In these cases, no pre-selection $\rm{S/B}$ threshold is applied. The single flares shown here are of three scenarios: a one-day flare (blue line), a ten-day flare (red lines), and a thirty-day flare (black line). The multiple flares are shown as dashed and dotted lines. All the results have been obtained for $\Delta T_{\rm{data}} = 3150$ days and 595 background events within a 12$^{\circ}$ x 12$^{\circ}$ sky region.
}
\label{fig:discovery}
\vspace{-0.1cm}
\end{figure*}

The discovery thresholds calculated for different methods are presented in Figure \ref{fig:discovery}. Shown are the multiplet (Braun) and the stacking methods, with and without application of the $S_4$ photon tag, for single and multiple flares. For the multiplet methods, no pre-selection involving the $\rm{S/B}$ signal-to-background threshold is applied, while the results for the stacking method are shown as a function of this parameter.
The results for flares of different durations are indicated by different colors, whereas multiple flares are distinguished using various line styles. The figure demonstrates the clear advantage of the stacking method when combined with the $S_4$ photon tag, which significantly reduces the minimum number of events required for flare(s) detection.

\vspace{-0.2cm}

\section{Conclusion}

We propose two improved methods for investigating direction-time clustering in UHECR (air-shower) data that have the potential to provide evidence for UHE photons, possibly produced by point sources (astrophysical flares). They can also contribute to refining the upper limits on the UHE photon flux and advancing the search for sources of neutral UHECRs. Preliminary estimates suggest that current photon flux limits could be improved by approximately a factor of two.
Among the two methods, the stacking method offers notable advantages: it is capable of identifying multiple flares of arbitrary temporal profile from a single source, and most importantly, when combined with the $S_4$ photon tag, the method requires remarkably few events to detect photon flares. 


We are developing a Python package, named \texttt{UHECluster} (available at \href{https://gitlab.com/uhecluster/uhecluster}{\nolinkurl{https://gitlab.com/uhecluster/uhecluster}}), to create a uniform modular platform for running all the algorithms discussed here as simple Python scripts. The package provides robustness and ease of use, as well as simplified debugging and algorithm upgrades for improved performance. The inclusion of the stacking method in this package is under development and will soon be incorporated to provide a more robust implementation of the direction-time clustering methods discussed here.

\section{Acknowledgments}

\small{The authors would like to thank the colleagues from the Pierre Auger Collaboration for all the fruitful discussions. 
We want to acknowledge the support in Poland from the National Science Centre, grants No. 2020/39/B/ST9 /01398 and 2022/45/B/ST9/02163 as well as from the  Ministry of Science and Higher Education, grant No. 2022/WK/12.
}


\bibliographystyle{apalike}
\bibliography{main}

\begin{thebibliography}{}

\bibitem[Borodai, 2025]{natalia}
Borodai, N. (2025).
\newblock Gamma/hadron discriminant variables in application to high-energy cosmic-ray air showers.
\newblock {\em PoS}, ICRC(2025)199.

\bibitem[Braun et~al., 2010]{BRAUN2010175}
Braun, J., Baker, M., Dumm, J., Finley, C., Karle, A., and Montaruli, T. (2010).
\newblock Time-dependent point source search methods in high energy neutrino astronomy.
\newblock {\em Astroparticle Physics}, 33(3):175--181.

\bibitem[Braun et~al., 2008]{BRAUN2008299}
Braun, J., Dumm, J., {De Palma}, F., Finley, C., Karle, A., and Montaruli, T. (2008).
\newblock Methods for point source analysis in high energy neutrino telescopes.
\newblock {\em Astroparticle Physics}, 29(4):299--305.

\bibitem[Góra et~al., 2011]{GORA2011201}
Góra, D., Bernardini, E., and {Cruz Silva}, A. (2011).
\newblock A method for untriggered time-dependent searches for multiple flares from neutrino point sources.
\newblock {\em Astroparticle Physics}, 35(4):201--210.

\bibitem[Ros et~al., 2011]{ROS2011140}
Ros, G., Supanitsky, A., Medina-Tanco, G., {del Peral}, L., D’Olivo, J., and {Rodríguez Frías}, M. (2011).
\newblock A new composition-sensitive parameter for ultra-high energy cosmic rays.
\newblock {\em Astroparticle Physics}, 35(3):140--151.

\bibitem[Stasielak et~al., 2025]{stasielak2025b}
Stasielak, J., Borodai, N., G\'ora, D., Niechciol, M., and P\k{e}kala, J. (2025).
\newblock {Improved method of searching for flares of neutral particles from point sources}.
\newblock {\em PoS}, UHECR2024:107.

\bibitem[Stasielak et~al., 2024]{stasielak2024}
Stasielak, J., Borodai, N., Góra, D., and Niechciol, M. (2024).
\newblock An improved method to search for flares from point sources of ultra-high-energy photons.
\newblock {\em Central European Astrophysical Bulletin}, to be published, arXiv:2412.13804v1.

\bibitem[{The Pierre Auger Collaboration}, 2022]{universe8110579}
{The Pierre Auger Collaboration} (2022).
\newblock Searches for ultra-high-energy photons at the {P}ierre {A}uger {O}bservatory.
\newblock {\em Universe}, 8(11).

\end{thebibliography}

\end{document}